\documentclass[epj]{webofc}
\usepackage[varg]{txfonts}   % Web of Conferences font
\wocname{EPJ Web of Conferences}
\woctitle{ICNFP 2017}
\newcommand{\beq}{\begin{equation}}
\newcommand{\eeq}{\end{equation}}
\newcommand{\bea}{\begin{array}}
\newcommand{\eea}{\end{array}}
\newcommand{\beqa}{\begin{eqnarray}}
\newcommand{\eeqa}{\end{eqnarray}}

\begin{document}
\selectlanguage{english}
\title{Lattice QCD at finite baryon density using analytic continuation}

%
% subtitle (optional, strongly discouraged)
%
%%%\subtitle{Do you have a subtitle?\\ If so, write it here}

\author{V.~G. Bornyakov\inst{1,2,3}\fnsep\thanks{\email{vitaly.bornyakov@ihep.ru}} \and
        D.~L. Boyda\inst{2,4} \and
        V.~A. Goy\inst{2} \and
        H.~Iida\inst{2,5} \and
        A.~V.~Molochkov\inst{2} \and
        Atsushi Nakamura\inst{2,5,6} \and
        A.~A. Nikolaev\inst{2,3} \and
        V.~I. Zakharov\inst{2,3} \and
        M.~Wakayama\inst{2,5}
}

\institute{NRC Kurchatov Institute - IHEP, 142281 Protvino, Russia,
\and
           School of Biomedicine, Far Eastern Federal University, 690950 Vladivostok, Russia,
\and
           NRC Kurchatov Institute - ITEP, 117218 Moscow, Russia,
\and
           School of Natural Sciences, Far Eastern Federal University, 690950 Vladivostok, Russia,
\and
           Theoretical Research Division, Nishina Center, RIKEN, Wako 351-0198, Japan,
\and
           Research Center for Nuclear Physics (RCNP), Osaka University, Ibaraki, Osaka, 567-0047, Japan
}

\abstract{
  We simulate lattice QCD with two flavors of Wilson fermions at imaginary baryon chemical potential.
Results for the baryon number density computed in the confining and deconfining phases at imaginary baryon chemical potential are used to determine the baryon number density and higher cumulants at the real chemical potential via analytical continuation.
}
\maketitle

\section{Introduction}
%---------------------
\label{sec:introduction}

Recent results of heavy ion collision experiments at RHIC \cite{Adams:2005dq} and  LHC  \cite{Aamodt:2008zz}
shed some light on properties of the quark
gluon plasma and the position of the transition line in the baryon density - temperature plane.
New experiments will be carried out at FAIR (GSI) and NICA (JINR).
To explore the phase diagram theoretically it is necessary to make computations in QCD at finite temperature and
finite baryon chemical potential. For finite temperature and zero chemical potential lattice QCD is the only ab-initio method available
and many results had been obtained. However, for finite baryon density lattice QCD faces the so-called
complex action problem (or sign problem).
Various proposals exist to solve this problem see, e.g. reviews~\cite{Muroya:2003qs,Philipsen:2005mj,deForcrand:2010ys} and yet it is still
very hard to get reliable results at $\mu_B/T>1$.
Here we consider the analytical continuation from imaginary chemical potential.

%%%%%%%%%%%%
The fermion determinant at nonzero baryon chemical potential $\mu_B$, $\det\Delta(\mu_B)$, is in general not real.
This makes impossible to apply standard Monte Carlo techniques to computations with the partition function
\beq
Z_{GC}(\mu_q,T,V) = \int \mathcal{D}U (\det\Delta(\mu_q))^{N_f} e^{-S_G},
\label{Eq:PathIntegral}
\eeq
where $S_G$ is a gauge field action, $\mu_q=\mu_B/3$ is quark chemical potential,   $T=1/(aN_t)$ is temperature, $V=(aN_s)^3$ is volume, $a$ is lattice spacing, $N_t, N_s$ - number of lattice sites in time and space directions.
%%%%%%%%%%%%%%%%

It is known that the standard Monte Carlo simulations are possible for  the grand canonical partition function
$Z_{GC}(\theta,T,V)$ for imaginary chemical potential $\mu_q=i\mu_{qI} \equiv iT\theta$.
since the fermionic determinant is real for imaginary $\mu_q$.

\begin{figure}[htb]
\centering
\includegraphics[width=0.35\textwidth,angle=0]{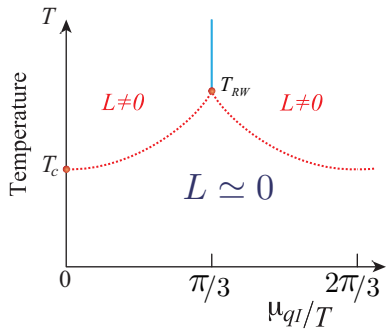}%
\vspace{0cm}
\caption{Schematical figure of Roberge-Weiss phase structure in the pure imaginary chemical
potential regions.}
\label{RW_ph_d}
\end{figure}

The QCD partition function  $Z_{GC}$ is a periodic function of $\theta$: $Z_{GC}(\theta) = Z_{GC}(\theta+2\pi/3)$.
This symmetry is called Roberge-Weiss symmetry \cite{Roberge:1986mm}.
%As a consequence of this periodicity
%the canonical partition functions $Z_C(n,T,V)$  are nonzero only for $n=3k$.
 QCD possesses a rich phase structure at nonzero $\theta$, which depends on the number of
flavors $N_f$ and the quark mass $m$. This phase structure is shown in
Fig.~\ref{RW_ph_d}.  $T_c$ is the confinement/deconfinement crossover temperature at
zero chemical potential.  The line $(T \ge T_{RW},\mu_I/T=\pi/3)$ indicates the
first order phase transition.  On the curve between $T_c$
and $T_{RW}$, the transition is expected to change from the crossover to the first order
for small and large quark masses, see e.g. \cite{Bonati:2014kpa}.
 Quark number density $n_q$ for $N_f$ degenerate quark flavours is defined by the following equation:
\beq
\frac{n_{q}}{T^{3}} = \frac{1}{VT^{2}}\frac{\partial}{\partial \mu_q}\ln
Z_{GC}
=\frac{N_{f}N_{t}^{3}}{N_s^3 Z_{GC}} \int \mathcal{D}U e^{-S_G} (\det\Delta(\mu_q))^{N_f}
\mathrm{tr}\left[\Delta^{-1}\frac{\partial \Delta}{\partial \mu_q/T}\right].
\label{density1}
\eeq
It can be computed numerically for imaginary chemical potential. Note, that for the imaginary chemical potential $n_q$ is also purely imaginary: $n_q = i n_{qI}$.

In this work
we fitted $n_{qI}/T^3$ to theoretically motivated functions
of $\mu_{qI}$. It is known that the density of noninteracting  quark gas is described  by
\beq
n_q/T^3 = N_f \Bigl ( \frac{\mu_q}{T}  + \frac{1}{\pi^2} \Bigl (\frac{\mu_q}{T} \Bigr )^3 \Bigr ).
\eeq
We thus fit the data for $n_{qI}$ to an odd power polynomial of $\theta$
\beq
n_{qI}(\theta)/T^3 = \sum_{n=1}^{n_{max}} a_{2n-1} \theta^{2n-1}\,,
\label{eq_fit_polyn}
\eeq
in the deconfining phase at temperature $T > T_{RW}$. This type of the fit was also used in Refs.~\cite{DElia:2009pdy,Takahashi:2014rta,Gunther:2016vcp,Bornyakov:2016wld}.
%and Ref.~\cite{Gunther:2016vcp}.

In the confining phase (below $T_c$) the hadron resonance gas model provides
good description of the chemical potential dependence of thermodynamic observables
\cite{Karsch:2003zq}.
Thus it is reasonable to fit the density to a Fourier expansion
\beq
n_{qI}(\theta)/T^3 = \sum_{n=1}^{n_{max}} f_{3n} \sin(3n \theta)
\label{eq_fit_fourier}
\eeq
Again this type of the fit was  used in Refs.~\cite{DElia:2009pdy,Takahashi:2014rta} and conclusion was made that it works well.
We use both types of the fitting function in the deconfining phase at $T_c < T < T_{RW}$.

   We made simulations of the lattice QCD with $N_f=2$ clover improved Wilson quarks and Iwasaki improved gauge field action. The more detailed definition of the lattice action can be found in Ref.~\cite{Bornyakov:2016wld}.
The simulations were made on $16^3 \times 4$ lattices. We obtained results at temperatures $T/T_c=1.35, 1.20$, 1.08, and 1.035 in the deconfinement phase  and $0.99, 0.93, 0.84$ in the confinement phase along the line of constant physics with
 $m_{\pi}/m_{\rho}=0.8$. We also present here our preliminary results for smaller quark mass with $m_{\pi}/m_{\rho}=0.65$  At this quark mass the simulations were made at $T/T_c =  1.32, 1.18, 1.07, 1.00, 0.94, 0.86$.
 The parameters of the action, including $c_{SW}$ value were borrowed from the WHOT-QCD collaboration paper~\cite{Ejiri:2009hq}. We compute the number density on samples of $N_{conf}$ configurations with $N_{conf}=1800$ or 3800, using every 10-th trajectory produced with Hybrid Monte Carlo algorithm.

\begin{figure}[t]
\centering
\includegraphics[width=7cm,clip,angle=270]{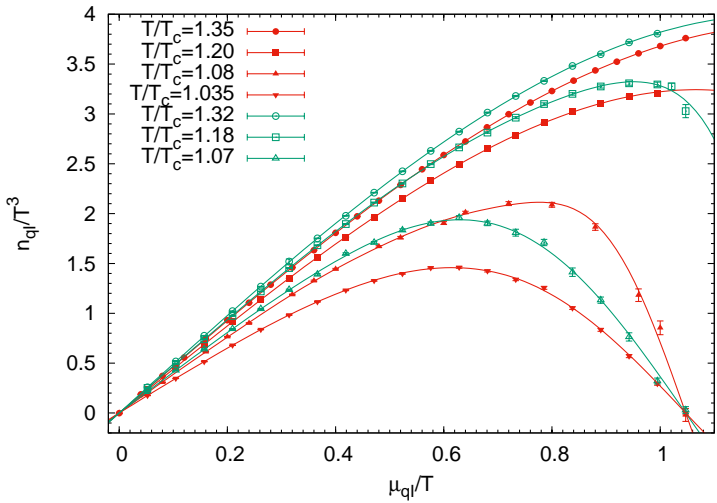}
\caption{The quark number density $n_{qI}$ in the deconfinement phase at two quark masses: $m_\pi/m_\rho = 0.8$ (filled symbols) and  $m_\pi/m_\rho = 0.65$ (empty symbols).}
\label{fig-2}       % Give a unique label
\end{figure}

\section{Quark number density}
\label{section-QM}

\begin{figure}[ht]
\centering
\includegraphics[width=7cm,clip,angle=270]{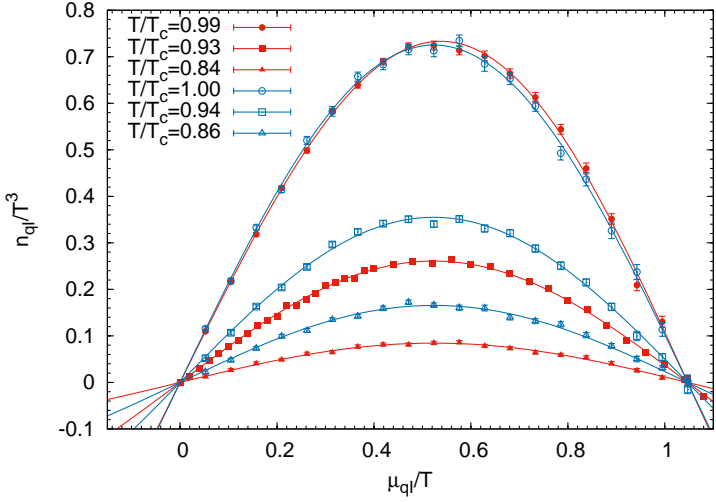}
\caption{The quark number density $n_{qI}$ in the confinement phase at two quark masses: $m_\pi/m_\rho = 0.8$ (filled symbols) and  $m_\pi/m_\rho = 0.65$ (empty symbols).}
\label{fig-3}       % Give a unique label
\end{figure}

In this section we compare results for the quark number density obtained at the imaginary chemical potential for two values of the
quark mass. In Figure~\ref{fig-2} the data for the deconfinement phase are shown. One can see that at small values of $\mu_{qI}/T$
the number density for two quark masses differ only slightly for comparable values $T/T_c$. At the same time effects of the quark mass decreasing are quite visible in the range $\mu_{qI}/T > 0.8$.  As can be seen from Figure~\ref{fig-3} in the confinement phase the differences between results for two quark masses might be mostly due to differences in the $T/T_c$ values.

\begin{figure}[ht]
\centering
\includegraphics[width=7cm,clip,angle=270]{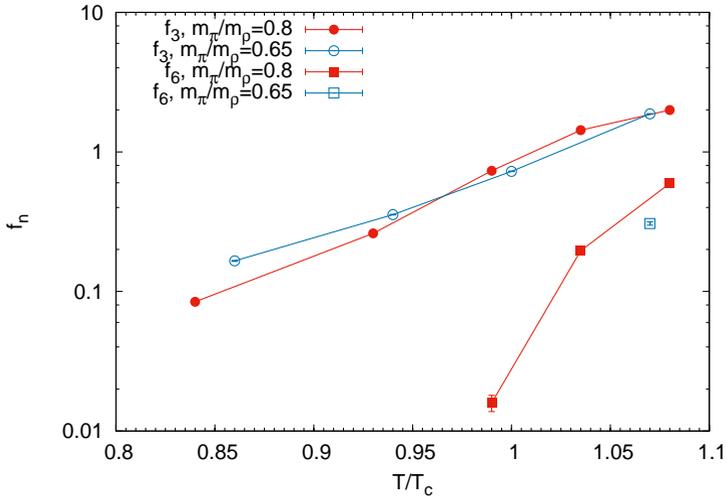}
\caption{The virial coefficients  $f_3$ and $f_6$ for $m_\pi/m_\rho = 0.8$ (filled symbols) and  $m_\pi/m_\rho = 0.65$ (empty symbols).}
\label{fig-4}       % Give a unique label
\end{figure}

This assumption is supported by comparison of the virial coefficients $f_n$ depicted in Figure~\ref{fig-4}. Note logarithmic scale for Y-axes in this Figure. From this Figure one can see an exponential decrease of $f_n$ with decreasing temperature. There is an indication of slower decrease for lower quark mass. Comparing the slopes for $f_3$ and $f_6$ we conclude that it is steeper for $f_6$.

\begin{figure}[th]
\centering
\includegraphics[width=7cm,clip,angle=270]{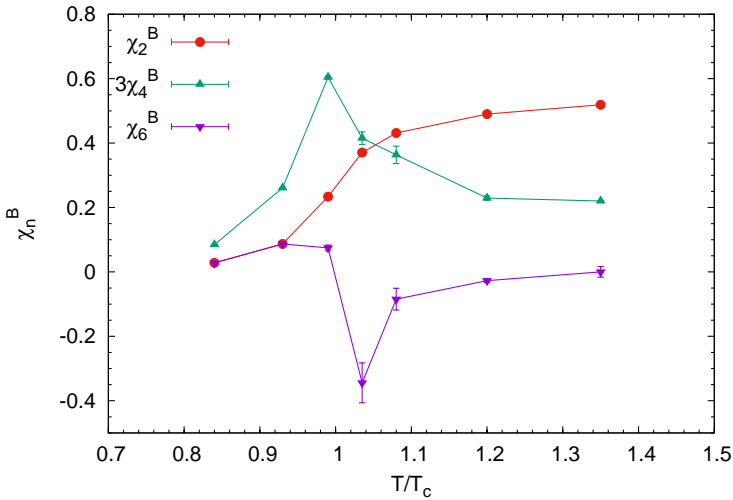}
\caption{The generalized susceptibilities for $m_\pi/m_\rho = 0.8$ .}
\label{fig-5}       % Give a unique label
\end{figure}

\section{Taylor expansion coefficients}

The Taylor expansion coefficients for the pressure are introduced as follows:
\beq
\Delta P(T,\mu_B) = \sum_{k=1}^{\infty} P_{2k}(T)~ \mu_B^{2k} = \sum_{k=1}^{\infty} \frac{1}{(2k)!}\chi_{2k}(T)~ \mu_B^{2k} \,,
\eeq
where $\Delta P(T,\mu_B) = P(T,\mu_B) - P(T,0)$, $P_{2k}$ are Taylor expansion coefficients, $\chi_{2k}$ are called generalized susceptibilities.
Before we discuss the Taylor expansion coefficients few comments about the fits follow.
The fits to function eq.~(\ref{eq_fit_fourier}) are very stable with respect to change of the fitting range. This was observed for
$T/T_c=0.84, 0.93$. Since the statistical error for the coefficient $f_3$ is small at low temperatures we obtain the Taylor coefficients $a_k$ with low error even for high values of $k$. We should note that there is a source of uncertainty which we cannot estimate reliably. This is  the contribution from the higher terms in the Fourier decomposition  eq.~(\ref{eq_fit_fourier}). Our data indicate that for low temperatures
$f_6$ is at least factor 100 smaller than $f_3$. This implies that for Taylor coefficients $a_1$ and $a_3$ the contribution from the second term in eq.~(\ref{eq_fit_fourier}) should be small while starting from $a_5$ this contribution might be substantial.

\begin{figure}[ht]
\centering
\includegraphics[width=7cm,clip,angle=270]{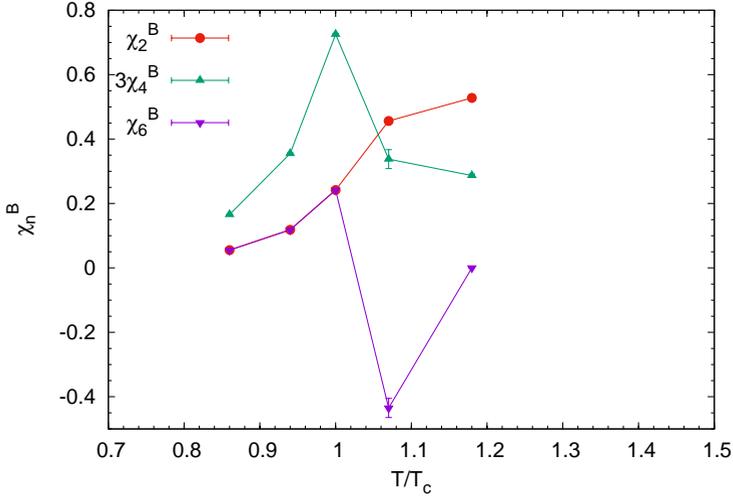}
\caption{The generalized susceptibilities for  $m_\pi/m_\rho = 0.65$ .}
\label{fig-6}       % Give a unique label
\end{figure}

\begin{figure}[ht]
\centering
\includegraphics[width=7cm,clip,angle=270]{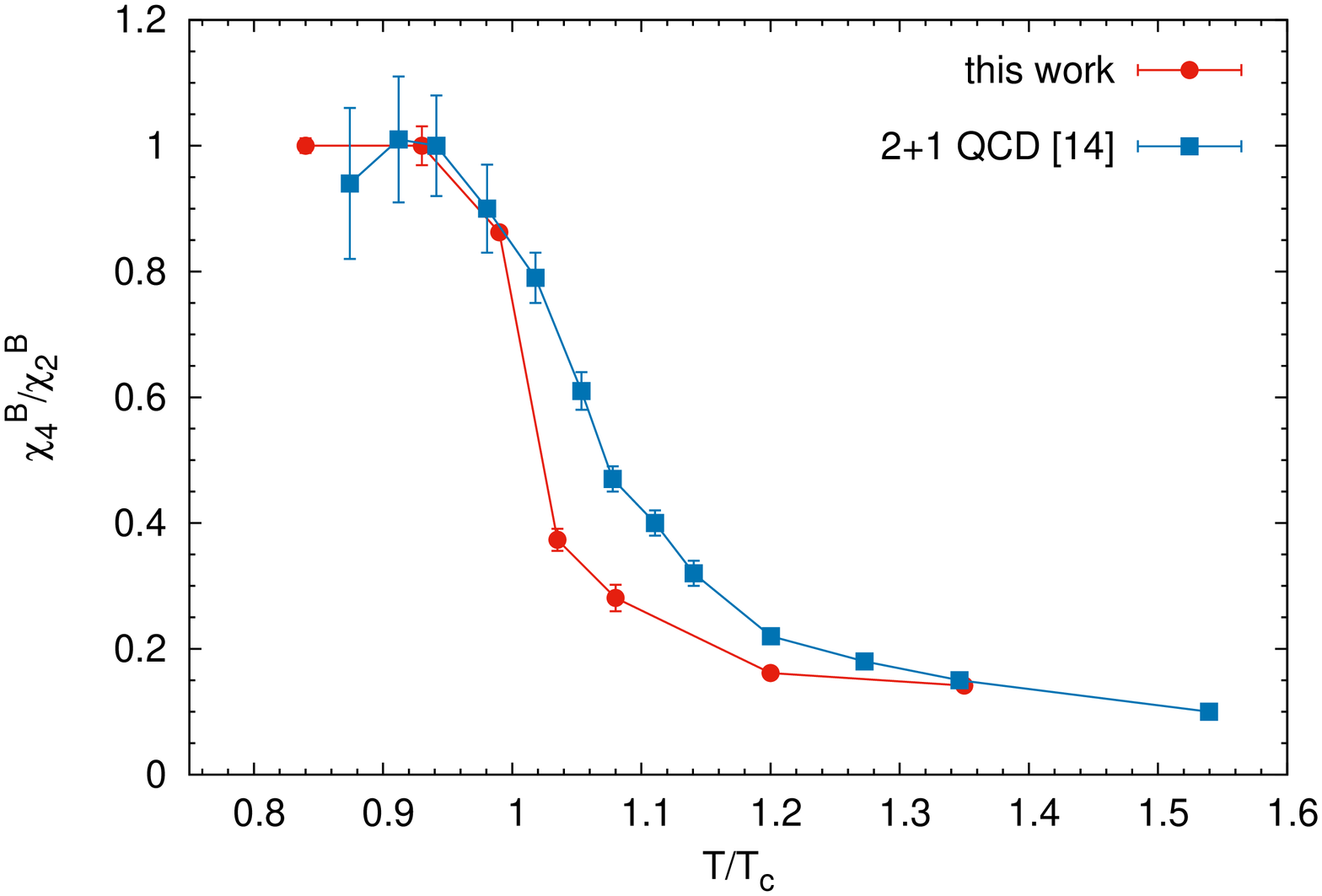}
\caption{The ratio $\chi_4^B/\chi_2^B$ at $m_\pi/m_\rho = 0.8$ in comparison with results from [14].}
\label{fig-7}       % Give a unique label
\end{figure}

\begin{figure}[ht]
\centering
\includegraphics[width=7cm,clip,angle=270]{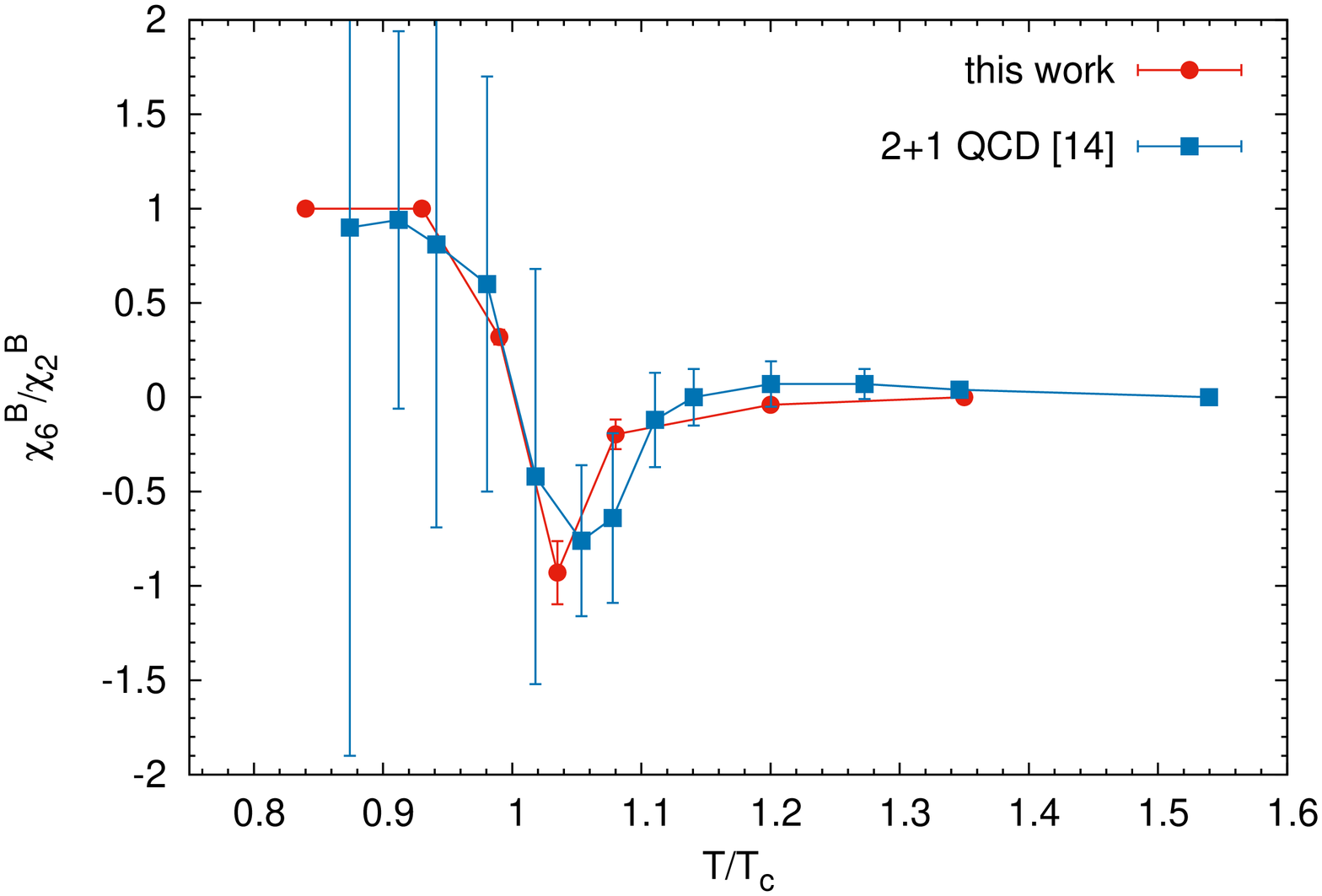}
\caption{The ratio $\chi_6^B/\chi_2^B$ at $m_\pi/m_\rho = 0.8$ in comparison with results from [14].}
\label{fig-8}       % Give a unique label
\end{figure}

For consistency check of our results we also applied polynomial fit at low temperatures. The polynomial fit was applied for restricted range of $\mu_{qI}$ values. We obtained results compatible with respective Taylor coefficients $a_k, k=1, 3, 5$ within error bars for $T/T_c=0.84, 0.93$. For $T/T_c=1.035$ and 1.08 the results for $a_k$ obtained with two kind of fits are in agreement for $k=1$ only. We then used $a_3$ and $a_5$ values obtained from the polynomial fit. For the lighter quark mass we made similar computations.

The generalized susceptibilities $\chi_n$ for $n=2, 4, 6$ which are proportional to the Taylor coefficients $P_n$ are presented in
Figure~\ref{fig-5} for   $m_\pi/m_\rho = 0.8$  and  in Figure~\ref{fig-6} for   $m_\pi/m_\rho = 0.65$. We should note that our results for $P_2$ and $P_4$  are in agreement within error bars with results of Ref.~\cite{Ejiri:2009hq} where direct computation of the Taylor coefficients was done. But our error bars are substantially lower.

The Taylor coefficients were recently computed for the physical quark masses on lattices with small lattice spacing (and even in the continuum limit) in Refs.~\cite{Gunther:2016vcp} and  \cite{Bazavov:2017dus}. In Ref.~\cite{Gunther:2016vcp} the analytical continuation was used
while in Ref~\cite{Bazavov:2017dus} direct method was employed. Results of these two computations were found to be in a good agreement \cite{Bazavov:2017dus}. Our results presented in Figure~\ref{fig-5} and Figure~\ref{fig-6} are in good qualitative  agreement with
results of Refs.~\cite{Gunther:2016vcp,Bazavov:2017dus} for all three susceptibilities. Quantitatively our results at $T>T_c$ are substantially  higher what should be expected from the large lattice spacing effects estimated in  Ref.~\cite{Ejiri:2009hq}.

In  Figure~\ref{fig-7} and Figure~\ref{fig-8} we show the ratios $\chi_4^B/\chi_2^B$ and $\chi_6^B/\chi_2^B$, respectively.
One can see that for these ratios our results are in very good agreement with results of Ref~\cite{Bazavov:2017dus} taken from
their Figure~3. Substantial difference is observed only for $\chi_4^B/\chi_2^B$ at  $1 < T/T_c < 1.1$. This agreement indicates that
the finite lattice spacing effects are substantially cancelled in the ratios of the susceptibilities.

\section{Conclusions}
We computed the baryon number density and generalized susceptibilities $\chi_n$ in the lattice QCD with two flavors of Wilson fermions
on $16^3 \times 4$ lattices using analytical continuation. Comparing results for two quark masses we found that they do not differ substantially. Comparison of our results for the ratios $\chi_4^B/\chi_2^B$ and $\chi_6^B/\chi_2^B$ with results of Ref~\cite{Bazavov:2017dus}
where simulations were done at the physical quark masses and small lattice spacing we found surprising agreement. This agreement indicates
that our recent results \cite{Boyda:2017dyo} showing agreement of cumulant ratios  computed on the lattice with respective experimental results
are not accidental.
\vspace{0.5cm}

\noindent
{\bf Acknowledgments} \\
This work was completed due to support by
RSF grant under contract 15-12-20008. Computer simulations were performed on the FEFU GPU cluster Vostok-1 and  MSU 'Lomonosov' supercomputer.

\end{document}